\documentclass[journal,comsoc]{IEEEtran}

\usepackage[T1]{fontenc}		
\usepackage[utf8]{inputenc}     

\usepackage[cmex10]{amsmath}
\usepackage{siunitx}
\usepackage{cite}

\usepackage{graphicx}
\usepackage{epstopdf}
\usepackage{multirow}
\usepackage{booktabs}

\ifCLASSOPTIONcompsoc
  \usepackage[caption=false,font=normalsize,labelfont=sf,textfont=sf]{subfig}
\else
  \usepackage[caption=false,font=footnotesize]{subfig}
\fi

\usepackage{url}
\usepackage{hyperref}            
\hypersetup{
    linktocpage,
    colorlinks,
    citecolor=black,
    filecolor=black,
    linkcolor=black,
    urlcolor=black
 } 

\title{Estimation of Received Signal Strength Distribution for Smart Meters with Biased Measurement Data Set}
\author{%
	\IEEEauthorblockN{%
    Mathias Rønholt Kielgast\IEEEauthorrefmark{1},~\IEEEmembership{Student Member,~IEEE,}
	Anders Charly Rasmussen\IEEEauthorrefmark{1},~\IEEEmembership{Student Member,~IEEE,}\\
	Mathias Hjorth Laursen\IEEEauthorrefmark{1},~\IEEEmembership{Student Member,~IEEE,}
    Jimmy Jessen Nielsen\IEEEauthorrefmark{1},~\IEEEmembership{Member,~IEEE,} \\
	Petar Popovski\IEEEauthorrefmark{1},~\IEEEmembership{Fellow,~IEEE,} 
	and Rasmus Krigslund\IEEEauthorrefmark{2},~\IEEEmembership{Member,~IEEE.}%
	}

	\IEEEauthorblockA{%
		\IEEEauthorrefmark{1}Department of Electronic Systems, Aalborg University, 9220 Aalborg, Denmark\\ 
    \IEEEauthorrefmark{2}Kamstrup A/S, 8660 Skanderborg, Denmark \\ 
		Emails: \it{\footnotesize \{mkielg11, arasm12, mhla12\}@student.aau.dk, \{jjn, petarp\}@es.aau.dk, rkl@kamstrup.dk}
	}}

\begin{document}
\bstctlcite{IEEEexample:BSTcontrol}		

\maketitle
\thispagestyle{plain}
\pagestyle{plain}

\begin{abstract}
This letter presents an experimental study and a novel modelling approach of the wireless channel of smart utility meters placed in basements or sculleries. The experimental data consist of signal strength measurements of consumption report packets. Since such packets are only registered if they can be decoded by the receiver, the part of the signal strength distribution that falls below the receiver sensitivity threshold is not observable. We combine a Rician fading model with a bias function that captures the cut-off in the observed signal strength measurements. Two sets of experimental data are analysed. It is shown that the proposed method offers an approximation of the distribution of the signal strength measurements that is better than a naïve Rician fitting. 
\end{abstract}


\begin{IEEEkeywords}
Channel characterization and modeling, Received signal strength measurements, Communication system planning.
\end{IEEEkeywords}

\section{Introduction} 		\label{sec:Intro}
\IEEEPARstart{U}{tility} companies are transitioning from simple standalone consumption meters towards an Advanced Metering Infrastructure (AMI), where wireless smart meters are read automatically, enabling new services and features in addition to simple metering and billing \cite{WhyAMI}.
Installing an AMI, with one or several concentrators covering the distribution area, can be a costly investment. Hence, some utilities choose an intermediate solution, where wireless smart meters are installed but read only from a vehicle driving through the neighborhood \cite{AMR}. This is referred to as drive-by reading. Data is collected whenever consumption measurements are required, e.g. mainly for billing purposes, and the utility is thus not harvesting all benefits of the smart meters. However, the system is in this way prepared for automatic reading from concentrators in order to exploit the full potential of an AMI solution.
A big challenge for the utility company is to determine the best location(s) for installing the concentrators, since the experienced wireless path loss is highly dependent on the smart meter location, topography, structural mass, and vegetation of a given site. While dedicated measurement campaigns or ray tracing simulations could in principle be used to find suitable concentrator locations, such approaches require too much time and resources in practice. Parametric models of the signal strength distribution such as the Rician model are more promising for concentrator placement, however the parameters must be determined for the specific scenario considered, which again requires a time-consuming measurement campaign.

In the case where a utility is upgrading from a drive-by solution, they will already have a database of measurements collected during drive-by readings that includes received signal strength (RSS) measurements. Unfortunately, the data is not suitable for directly parameterising for example a Rician model, since measurement packets whose received signal strength is below the receiver sensitivity threshold are not decodable and the dataset is therefore incomplete.
In this work, we propose a method for fitting the Rician distribution parameters to the sample data by estimating the unobservable segment of the signal strength distribution. 

Numerous methods for estimation of the Rician K-factor exist, both using maximum likelihood methods \cite{RicianML}, as well as moment based methods \cite{RicianMM,Biasestimationrician}; however, none of these methods take the measurement bias into account. 	
Instead, we use a least-square optimisation to fit the parameters of a cut-off Rician distribution to the observed data. Thereby, we reduce the root mean square error (RMSE) on the CDF compared to when the parameters are fitted to the observed data with a standard Rician distribution.
We demonstrate the proposed estimation method for two experimental data sets:
drive-by measurements and measurements from a fixed concentrator setup. The results indicate a clear similarity between the two measurement sets in terms of signal strength distributions.

\section{System Model} 		\label{sec:SystemModel}
The field measurements used for modelling the signal strength distribution are obtained from drive-by smart meter readings and concentrator smart meter readings. The drive-by data set is collected using a Kamstrup \textsc{read}y converter \cite{ready_conv}, which is connected to an Android device running the so-called \textsc{read}y app.
When configured for the drive-by readings, the smart meters transmit data every \SI{16}{\second}. 
On the other hand, for the scenario with a concentrator, the transmit interval is increased to \SI{96}{\second}, which enables a larger signal power without reducing battery life time of the smart meter, while still providing a suitable data resolution for the relevant uses.

\begin{table}[ht]
\centering
\renewcommand{\arraystretch}{1.1}
\caption{System Specifications}
\label{tab:spec}
\begin{tabular}{lcc}
\toprule
	\textbf{Specification}              & \textbf{Value}  	& \textbf{Unit} \\ \cmidrule{1-3}
	Modulation 							& BFSK 				& - 			\\
	Frequency                           & 868.95     		& MHz           \\
	Bandwidth                           & 200        		& kHz           \\
	Transmit power (drive-by)           & 8          		& dBm           \\
	Transmit power (fixed network)      & 14         		& dBm           \\
	Receiver sensitivity        		& $S$        		& dBm           \\
\bottomrule
\end{tabular}
\end{table}

\begin{figure}[b]
  \centering
  \includegraphics[width=0.38\textwidth]{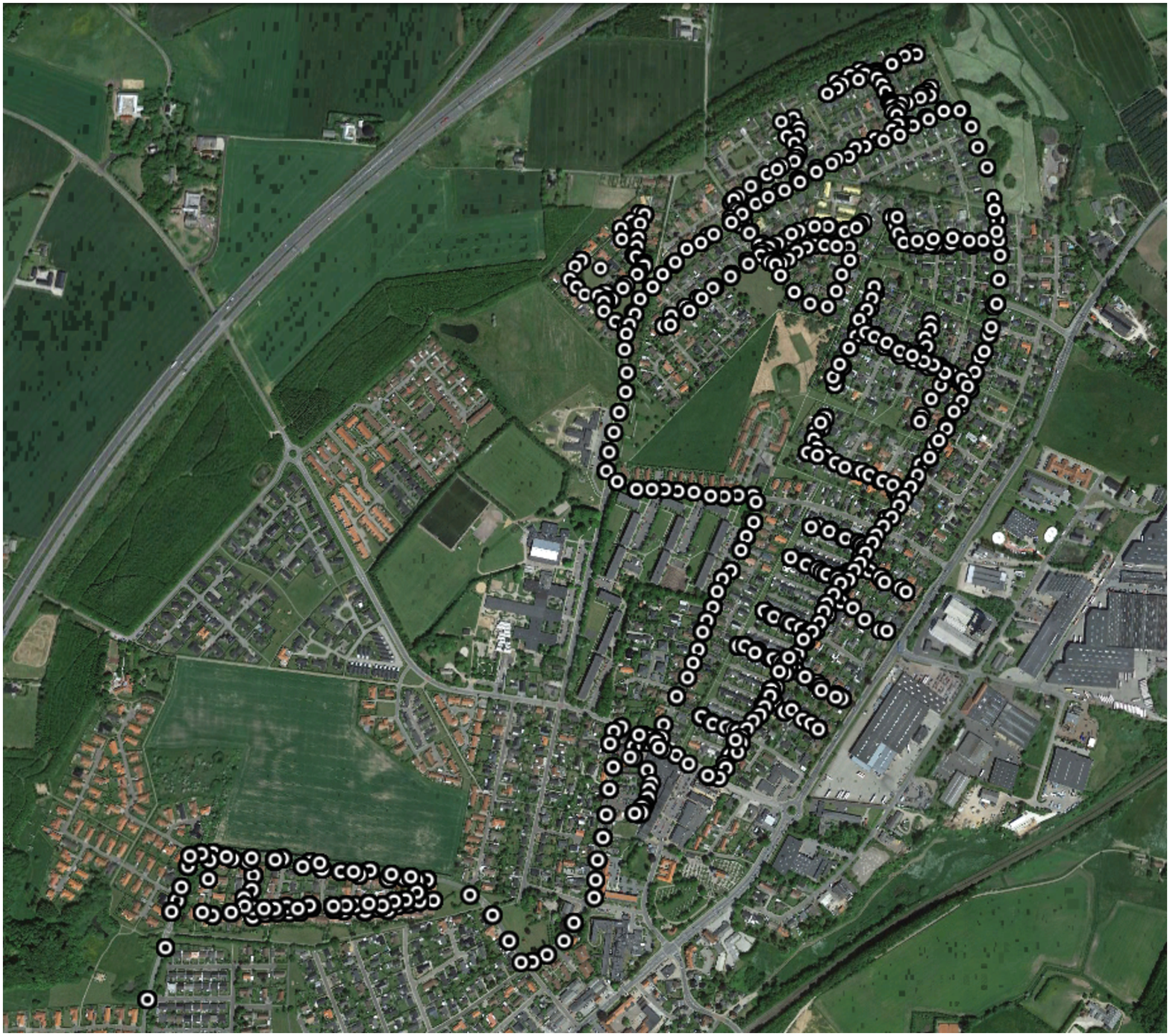}
  \caption{White dots are positions of the vehicle where data has been received. Map from \cite{GEarth}.}
  \label{fig:roads}
\end{figure}

The smart meters are mounted in houses, usually in sculleries or below ground level in basements. 
Hence, the signal can be attenuated to a high degree as discussed in \cite{TR45820}, meaning that the drive-by vehicle has to be within a reasonably short range to receive the signal, due to propagation loss in the wireless channel.
Fig. \ref{fig:roads} displays the position where the \textsc{read}y converter has received readings from smart meters. Only a single reading is received for each position, as the car is moving at all times. The measurement data for the concentrator scenario has been obtained from a different, but similar neighbourhood in terms of building types, density, vegetation, etc.

The transmissions are based on the wireless M-Bus standard (EN13757-4, C mode, defined in \cite{WMBUS}). The specifications of the system parameters are presented in Table \ref{tab:spec}.
\begin{figure}[t]
  \centering
  \includegraphics[width = 0.43\textwidth]{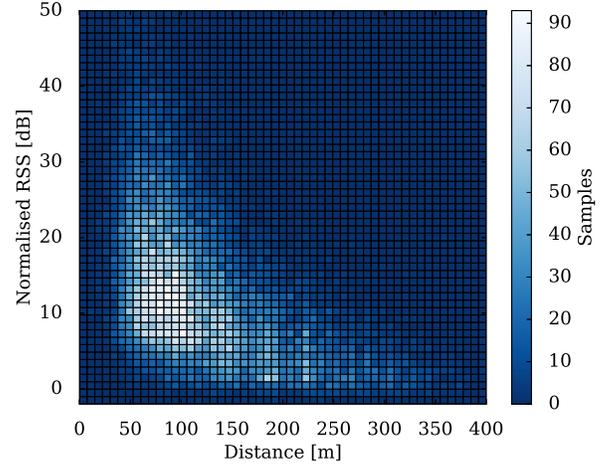}
  \caption{Frequency of sampled RSS as a function of the distance between receiver and the transmitting smart meter. Notice that the RSS is offset by the confidential sensitivity threshold $S$.}
  \label{fig:PLdata}
\end{figure}
The receiver sensitivity, $S$, is the point at which packets of 20 bytes are received with a success rate of \SI{20}{\percent}, and its value depends on the characteristics of the specific receiver device used. For the Kamstrup \textsc{read}y converters and concentrators, the exact value of $S$ is confidential. Hence, the data plotted in Fig. \ref{fig:PLdata} is normalised with regards to the converter sensitivity, $S$. It can be noted that practically no packets are received below the sensitivity boundary.


\section{Methods}\label{sec:Methods}
Empirical path loss models have been tested to find correlation between their predictions and the measured data. However, due to the large variance of the data, especially for smaller distances as can be seen in Fig. \ref{fig:PLdata}, this deterministic approach of using models to predict the average power is not sufficient. The first step, which we tackle in this paper, is to estimate the signal strength distribution. 

As discussed in \cite{MobileRadio}, the envelope of the received signal for a specific distance will be Rician distributed somewhere between pure line-of-sight (LOS) and Rayleigh distribution. Due to this, it is desired to estimate the distribution of RSS for the specific environment, with a higher accuracy than the established empirical models. This is done by curve fitting the cumulative density function (CDF) of the dataset in Fig. \ref{fig:PLdata} to a Rician distribution. For this, the Rician probability density function (PDF) \cite{MobileRadio} is used:
\begin{equation} \label{eq:Rician}
f_\text{Rician}(r) = \frac{2 r K}{r_\text{s}^2} \exp\left(-\frac{K}{r_\text{s}^2}(r^2+r_\text{s}^2)\right) \cdot I_0\left(\frac{2 r K}{r_\text{s}}\right) \,.
\end{equation}
Here, the distribution of the signal strength, $r$, is defined by the two parameters $K=10^{K_\text{dB}/10}$ and $r_\text{s}$.
$K_\text{dB}$ denotes the ratio between the power of the dominant component ($r_\text{s}$, usually LOS) and the power of the multipath components in dB.

\begin{figure*}[t]
	\centering
	\subfloat[Drive-by     \label{fig:DriveByData}]{\includegraphics[width=0.42\linewidth]{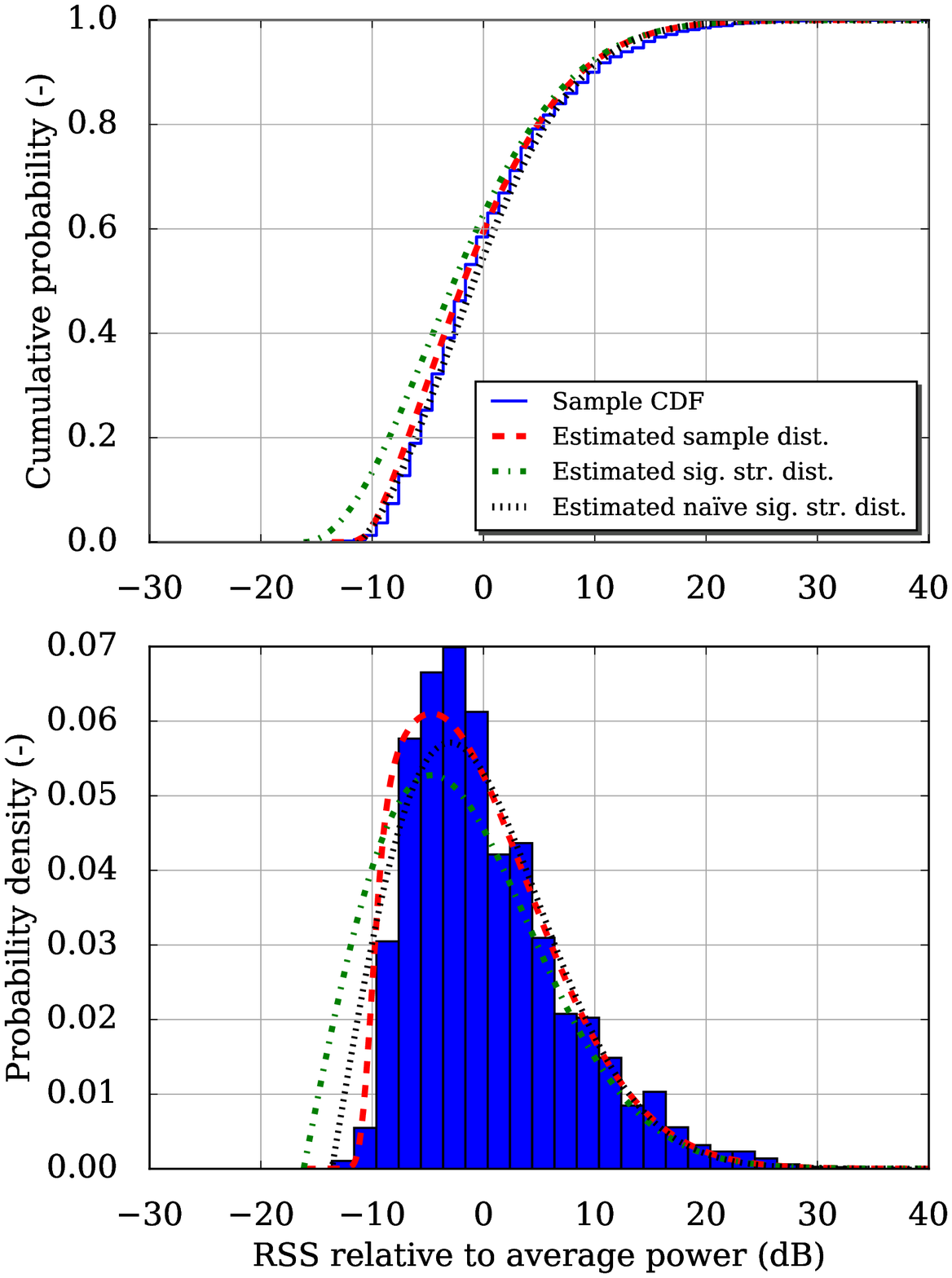}}
	\hspace*{12mm}
	\subfloat[Concentrator \label{fig:ConData}]    {\includegraphics[width=0.42\linewidth]{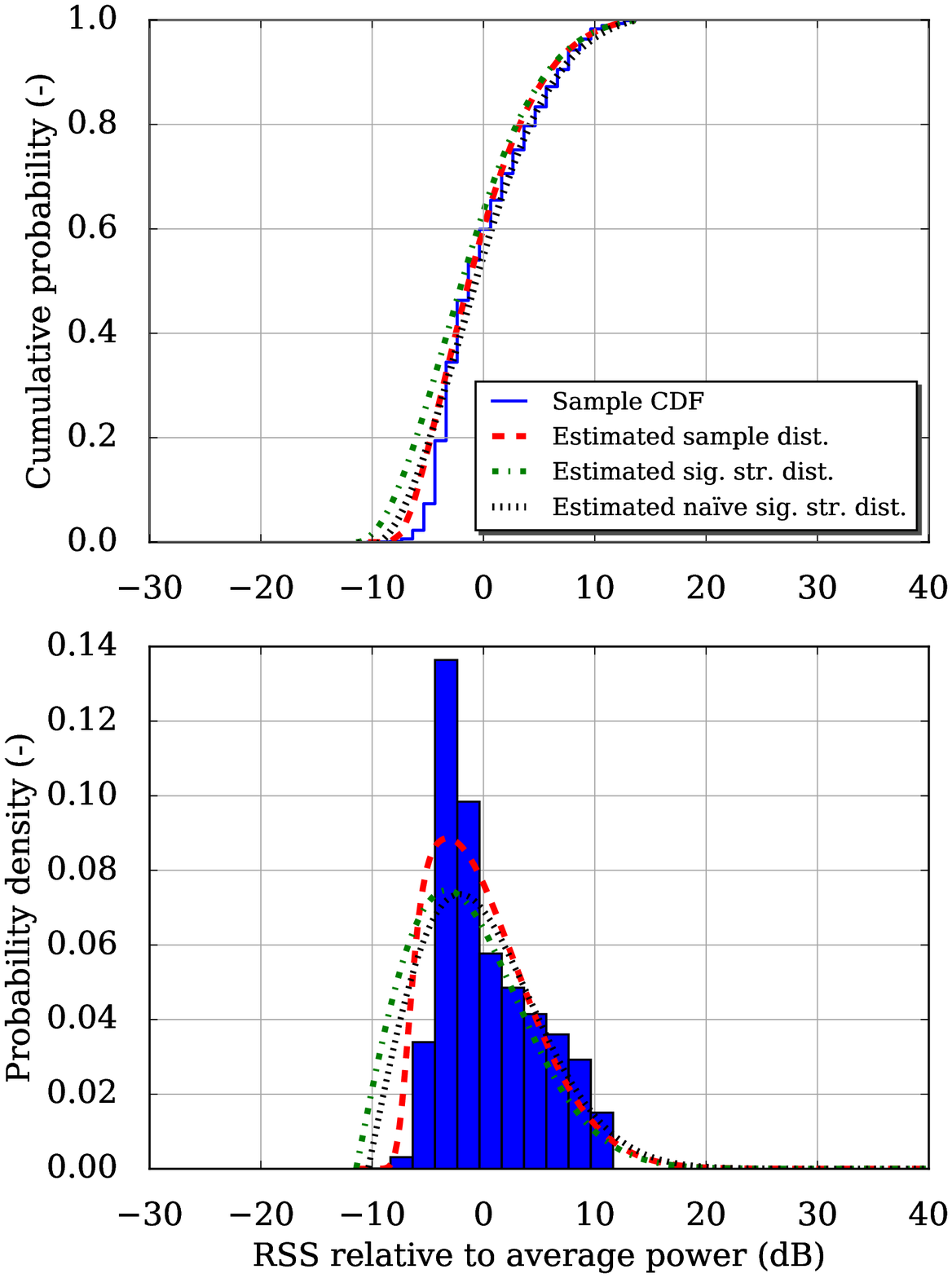}}
	\caption{PDF and CDF of Rician distribution fitted to measurement sample data. Note that the $x$-axis is normalised with regards to the mean of the sampled signal strengths, which is different for the concentrator data and the drive-by data.}
	\label{fig:complete_cdf_est}
\end{figure*}

Even though it may be possible to fit this distribution to the measurement data (black curve in Fig. \ref{fig:DriveByData}), it would be improper to assume that it depicts the actual signal strength distribution. This is due to the cut-off of the data. To describe this cut-off or bias of the data, the following is defined:
\begin{equation*} 
  \begin{aligned}
    f(r): \hspace*{3mm}&\text{signal strength distribution \hspace*{2mm}(PDF),} \\
    f_\text{sample}(r): \hspace*{3mm}&\text{sample distribution \hspace*{13mm}(PDF),} \\
    w(r):   \hspace*{3mm}&\text{bias function.}
  \end{aligned}
\end{equation*}
The distribution we want to estimate is the unobservable signal strength distribution, $f(r)$. To estimate this, we describe the observable sample distribution, $f_\text{sample}(r)$, as a function of the unobservable $f(r)$ and the bias function, $w(r)$, as:
\begin{equation}\label{eq:sampledist}
  f_\text{sample}(r) = \frac{w(r)f(r-r_0)}{\displaystyle\int_{-\infty}^{\infty}w(r')f(r')\;\mathrm{d}r'} \,.
\end{equation}
Note that the denominator normalises the resulting density function so that it integrates to unity.
Also, as the distribution does not start in the origin, an additional parameter $r_0$ is introduced, which is an offset of the signal strength distribution.

The bias function, $w(r)$, is in our case the probability for successfully receiving a packet at a given signal-to-noise ratio (SNR). Since this probability is exactly $1-\mathit{FER}$, where $\mathit{FER}$ is the frame error rate, we will use $1-\mathit{FER}$ as the bias function. In the considered Kamstrup system, packets are uncoded, modulated with BFSK, and received with non-coherent detection. The first step in determining the FER is to determine the BER as~\cite{ComSysEng}:
\begin{equation}
	P_\mathrm{BFSK} = \frac{1}{2}\exp\left(-\frac{\varepsilon_b}{2N_0}\right)\,.\label{eq:p_bfsk}
\end{equation}
Here, $\frac{\varepsilon_b}{N_0}$ is the ``bitwise SNR'' that is a scaling of the SNR by the bit rate and bandwidth. The SNR is found using the RSS, the theoretical thermal noise floor, and the noise figure of the receiver. However, we note that the noise figure is confidential for the meters used in our experiments. If the packet contains $M$ BFSK symbols, then FER is determined as:
\begin{equation}
	\mathit{FER} = 1 - \alpha(1 - \beta \mathit{BER})^M \,,
\end{equation}
where $M$ is the amount of bits, $\alpha = [0,1]$ is the preamble detection probability and $\beta=[0,1]$ can be used to account for forward error correction (FEC).
Since in the considered case study we use BPSK without FEC ($\beta=1$) and we assume ideal preamble detection ($\alpha=1$), our bias function is:
\begin{equation}
  w(r) = 1 - \mathit{FER} = (1 - P_\mathrm{BFSK})^M \,,
\end{equation}
where $P_\mathrm{BFSK}$ depends on the RSS through $\frac{\varepsilon_b}{N_0}$. We consider a packet length of \SI{50}{bytes}, i.e. $M=400$. 

By assuming that the unobservable $f(r) \sim f_\text{Rician}(r)$, i.e. that it follows the Rician distribution in eq. \eqref{eq:Rician}, we need just to determine the values of the parameters $K$, $r_s$, and $r_0$, which results in the sample distribution that best fit the observed data.
For this, we apply a non-linear least square error minimisation method (Levenberg-Marquardt algorithm, discussed in \cite{MethodsLMA, TheLMA} and implemented in \cite{lmfit}) to fit the cumulative sum of the distribution function $f_\text{sample}(r)$ to the empirical CDF of the observed data in Fig. \ref{fig:PLdata}. Specifically, we find a set of values for the parameters $K$, $r_\text{s}$, and $r_0$ that minimise the error between the empirical CDF of the observed data and the distribution function in Eq. \eqref{eq:sampledist}. 
The found values of $K$, $r_\text{s}$, and $r_0$ characterise the best Rician approximation of the true unobservable distribution of signal strengths.

While we have presented the method above in the context of a Kamstrup system based on the Wireless M-Bus, the method can be applied to other range-limited systems, e.g. Wi-Fi systems, by simply inserting the scenario-relevant equations in eq. \eqref{eq:Rician} and eq. \eqref{eq:p_bfsk}. The underlying principle of estimating the unobservable signal strength distribution by matching eq. \eqref{eq:sampledist} to the measurements, would be essentially the same.

\section{Results and Discussion}	\label{sec:ResDisc}
For the numerical evaluation, we have used the drive-by and concentrator measurement data described in Sec. \ref{sec:SystemModel}. Specifically, we have extracted measurements from both data sets from the distance range $75 - 125$ m. Thereby, we reduce the impact of path-loss in the signal strength distribution of the measurement data set.

In addition to the results of the proposed biased Rician fitting method, we also consider baseline results that are created using a naïve Rician fitting that does not take into account the cut-off in the measurements. Notice that the results are anonymised in the sense that they are offset by the confidential receiver sensitivity, $S$.
In addition to the graphical comparisons in Fig. \ref{fig:complete_cdf_est}, we also present the resulting parameters of the distribution fittings in Table \ref{tab:ricean_fitting}.

Primarily, the results show that our proposed biased fitting method achieves a significantly lower RMSE than the naïve Rician fitting method. The RMSE is reduced by more than 40\% for both the drive-by and concentrator measurement data. This finding confirms that the proposed method does indeed improve the distribution fitting procedure, when dealing with data sets that are cut-off due to receiver sensitivity limitations. The reason for the remaining RMSE may be due to the fact that the number of measurements in the data set is insufficient to reach lower RMSE levels. The slightly non-smooth shape of the sample data also indicates this.

Secondly, a comparison of the resulting parameters for the drive-by and concentrator scenarios shows that the signal strength distributions differ to some degree. Especially, the shape-determining $K$ parameter of the biased Rician fitting differs by approximately $3$ dB for the two scenarios.

\begin{table}[htb]
\centering
\caption{Fitted Rician parameters. $r_0$ is relative to receiver sensitivity $S$.}
\begin{tabular}{clcccc}
\toprule
   &  & $K$ [dB] & $r_\text{s}$ [dB] & $r_0$ [dB]& RMSE     \\ \cmidrule{3-6}
\multirow{2}{*}{\textbf{Drive-by}} & Naïve  & $-34.08$ & $0.30$       & $-1.00$ & $0.018$ \\
                                  & Biased & $-33.45$ & $0.35$       & $-3.53$ & $0.010$ \\ 
\multicolumn{6}{l}{} \\[-1.4ex]
\multirow{2}{*}{\textbf{Concentrator}} & Naïve  & $-29.66$ & $0.38$       & $-1.00$ & $0.045$ \\
                                      & Biased & $-30.40$ & $0.35$       & $-2.08$ & $0.026$ \\
\bottomrule
\end{tabular}
\label{tab:ricean_fitting}
\end{table}

\section{Conclusion}				\label{sec:Conc}
We have proposed a method for estimating the signal strength distribution when only biased data sets are available. The estimation method applies a model of the bias that can be produced heuristically, theoretically, or by use of assumptions of the sampling method; the model is then used to take the biasing into account when doing parameter estimation. The estimation method is not limited to one specific probability distribution, but when applying the method, one must assume a distribution and decide which parameters to estimate; this enables the method to be utilised within a wide range of applications, where distributions are sought to be estimated from biased data sets. Specifically, we formulate the bias function to be used for Kamstrup smart meters data collection, which models the frame error probability of measurement packet sent with non-coherent BPSK modulation.

The method has been applied to estimate the parameters of a Rician distribution using signal strength measurements from a drive-by and a concentrator scenario that in both cases are biased due to the receiver front end sensitivity. 

We find that in both cases we are able to closely fit our biased Rician distribution to the measurement data compared to a naïve direct fitting of measurement data. We also find that the fitted Rician distributions for the two scenarios have different $K$ parameters, meaning that a model obtained from drive-by measurements cannot be immediately used to predict the signal strength conditions in a concentrator scenario.

Since the collected measurements have a clear relation to distance-dependent path-loss, a next step is to jointly estimate the signal strength distribution and path loss parameters.

\section*{Acknowledgement}
The authors would like to thank Benjamin Grosbois for his invaluable work and Assistant Professor Wei Fan for guidance. 
This work is partially funded by EU, under Grant agreement no. 619437. The SUNSEED project is a joint undertaking of 9 partner institutions and their contributions are acknowledged.

\bibliographystyle{IEEEtran}


\end{document}